# A multi-scale study to unravel the dehydration mechanism of hydrated salts

Anne C. Claude[1], Hannelore Derluyn[1,2], Jorik van de Groep[1], Noushine Shahidzadeh[1]*

[1] Institute of Physics, University of Amsterdam, Science Park 904, 1098 XH, Amsterdam, Netherlands.
[2] Universite de Pau et des Pays de l'Adour, CNRS, LFCR, DMEX, Pau, France
* Corresponding author. Email: n.shahidzadeh@uva.nl

## ABSTRACT

**Understanding the dehydration mechanism of hydrated salts remains fundamentally important in solid-state chemistry, as their behavior affect fields ranging from heat storage to heritage conservation and pharmaceutical crystallization. Combining Raman confocal microscopy, dynamic weight-loss measurements, SEM, and micro-CT, we show that dehydration kinetics of sodium sulfate decahydrate (mirabilite) unfold through two regimes: an initial nucleation-controlled phase, where atomic rearrangement drives two-dimensional lateral growth following an exponential law, and a subsequent phase-boundary-controlled regime, where growth advances into the crystal depth, limited by water-vacancy formation at the hydrated–dehydrated interface. The resulting thenardite product shows ~35% shrinkage and forms a porous, layered dual-porosity nanocrystalline structure, with relative humidity directly governing crystal size. Extending this analysis to other sulfate hydrates reveals that crystallographic symmetry changes between hydrate and anhydrous phases dictate surface morphology, linking microstructure to mechanism. This multiscale framework uncovers dynamics hidden from bulk measurements and suggests structural indicators could predict dehydration pathways in other hydrated salt families, and opening route toward designing thermal energy storage materials without exhaustive experimental screening.**

## INTRODUCTION

Hydrated salts are not only abundant on Earth but also in other planets such as Mars *(1, 2)*, providing crucial information about the planet's hydrogeologic history. In addition, these salts are crystalline compounds that contain water molecules incorporated into their crystal structure as an essential component. These water molecules, known as "water of crystallization" or "water of hydration", are present in definite proportions and are often necessary for the crystal to maintain its structure. Because of the water in their crystalline structure, these salts are also significant natural materials which have major impact in various fields such as heritage conservation, pharmaceutical crystallization and heat storage. Cyclic hydration-dehydration sulfate salts inside porous stone generates crystallization pressure and stress, a major driver of weathering and damage to monuments and artefacts. Hydrate/anhydrate transitions affect the stability, solubility, and bioavailability of drug crystals during processing and storage, making dehydration behavior essential to formulation control. Finally, salt hydrates are investigated as thermochemical materials and phase change materials for renewable energy storage. Salt hydrates have very high energy densities (1-3 GJ/m$^3$ (*3*)), by producing heat during the hydration and releasing heat during dehydration. Therefore, they can be used to store heat during a warm day (cooling), and to release it when it is cold at night (heating) (see equation 1).

$$MX\cdot nH_2O\,(s) + Q \rightleftharpoons MX\cdot mH_2O\,(s) + (n-m)\,H_2O\,(g) \quad (1)$$

However, long-term sustainable use of these materials remains limited due to technical challenges including broad transition temperatures, supercooling hysteresis, and cycling degradation (*i.e.* irreversible phase transformation). The mobility of the water molecules, *i.e.* the strength of water-crystal interactions (*4*–*7*), within a crystal structure can be assessed by examining how readily the water of hydration evaporates from the crystalline structure. For a given salt, the higher the hydrate, the lower the temperature at which it loses water. Therefore, successive dehydration steps require successively higher temperatures, and the evaporation of the last water is the most challenging since the change in enthalpy of the reaction increases with decreasing hydration number (equation 1).

In our previous work we have shown how the mobility within the crystal can give rise to a surprising softness of the crystals during deliquescence (*8*). Removing the water (dehydration) on the other hand, typically alters the crystalline structure and therefore poses a challenge for reversibility. Understanding the (de)hydration mechanism of hydrated salts is thus of fundamental importance to optimize the performances in terms of reversibility (efficiency and longer lifetimes) of these compounds for thermal energy storage applications. However, the underlying details of the dehydration process in particular remain poorly understood.

In the years 1990 to 2000, models from solid-state physics were used to describe dehydration kinetics and mechanisms of hydrated salts, with experiments carried out usually under dry nitrogen atmospheres (*9*). However, around 20 years later (*10*–*16*) some authors pointed out that these models are not suitable to describe the dehydration reaction since they are neither taking into account the water vapor generated throughout the process, nor the moisture present in the surrounding if the experiment is not performed under dry nitrogen.

Here, we provide novel insights into the dynamics of dehydration by performing a multiscale study of the dehydration of sodium sulfate decahydrate ($Na_2SO_4 \cdot 10H_2O$) − also known as mirabilite − and its transformation into thenardite ($Na_2SO_4$) at room temperature for different relative humidities (RH). This is a particularly relevant material to study on the one hand for a fundamental point of view and on the other hand for applications in thermal energy storage since its energy density is 2.4 $GJ.m^{-3}$ or 1.6 $MJ.kg^{-1}$ (*17*), highly superior to a Li-ion batteries (0.72-1.8 $GJ/m^3$ or 0.46-0.72 $MJ.kg^{-1}$) (*18*, *19*). From the fundamental study of hydration/dehydration, $Na_2SO_4$ is a salt that essentially breaks the "stepwise" picture altogether. Mirabilite has no thermally accessible intermediate hydrate. In mirabilite, all ten waters are coordinated to $Na^+$ in edge-sharing $Na(H_2O)_6$ octahedra, so there is no way to strip an outer shell while leaving a stable lower hydrate (*20*). Subsequently , the transition between the "hydration states" 10 → 0 is not disrupted by the intermediate formation of the metastable phase $Na_2SO_4 \cdot 7H_2O$ (*17*, *21*). By combining different high-resolution techniques including X-ray microtomography, scanning-electron microscopy (SEM), and confocal Raman micro-spectroscopy we investigate the dehydration mechanism both at micro and macroscale. From their combined findings, we show that the well-studied macroscale dehydration that is characterized by the "phase-boundary" model in fact relies on preliminary nucleation events that are only visible at the microscale.

First, we perform a systematic study of the macroscale dehydration of a single crystal of mirabilite using dynamic weight loss measurements at different RH and identify the creation of water vacancies as the rate-limiting step in the dehydration that follows the "phase-boundary" regime. Second, we investigate the microscale mechanism of the dehydration and observe the apparition and growth of circular sites of the anhydrous phase on the surface of the hydrated crystal that are indicative of a two-dimensional nucleation regime preceding the bulk growth of the dehydrated zones. Remarkably, this initial nucleation regime at the early stage of dehydration is obscured in macroscale experiments. Third, we study the morphology of the crystals after dehydration at different RH and observe that the RH has a significant impact on the nanoscale morphology of the resulting particles, which may impact the rehydration rate of the product at subsequent cycles. Finally, we compare the microstructure of other sulfate-based hydrated salts and conclude that the initiation of circular nucleation sites rather than linear cracks generally only occurs when the crystallographic structure between the hydrate and the anhydrous is different.

Overall, our findings provide a crucial insight in the micro-scale mechanisms underlying the dehydration process, highlighting the fundamental origin of structural phase changes that affect the irreversibility during rehydration.

## RESULTS

### Macroscale study: Dehydration of sodium sulfate decahydrate macrocrystals

First, mirabilite macro crystals are formed in bulk solution from a supersaturated salt solution by dissolving $Na_2SO_4$ (Sigma Aldrich; 99.8% purity) little by little while stirring in Millipore water ($\rho \approx 18.2 M\Omega.cm$) up to a concentration of 18.6 wt% or 1.6 $mol.kg^{-1}$ of solvent. Beyond this concentration the solution becomes turbid due to the precipitation of mirabilite crystals ($Na_2SO_4 \cdot 10H_2O$) from the supersaturated solution (concentration at saturation being $m_s$ = 1.4 mol. $kg^{-1}$ of solvent). The stirring is then stopped and the bottle kept at rest for 1-2 days at room temperature at $T$ = 21°C. Precipitated mirabilite macro crystals in equilibrium are subsequently removed from the saturated salt solution for the dehydration experiments. Macro crystals of mirabilite of similar mass and geometry collected from the bottle (saturated salt solution) are let to dry one by one using dynamic weight loss measurements with a precision sartorius balance (precision 0.001 g) under controlled environment of RH and T. Practically, the dehydration of mirabilite is humidity-controlled rather than temperature-controlled: the thenardite–mirabilite equilibrium sits at 76.4 % RH at 20 °C. Therefore, this reaction of dehydration is known to be spontaneous at room temperature and $RH < RH_{eq} \approx 76$ % (*22*). This transition can be observed by naked eye at room temperature; the latter gradually turns from an initial transparent state to a white and opaque appearance upon dehydration, while retaining the overall shape of the initial mirabilite crystal (Fig. 1A). A SEM image of the dried crystal (Fig. 1B) reveals that the dehydrated crystal is in fact a microporous network comprised of a nanocrystal assembly which already suggests a structural phase transition within the bulk of the crystal. (*8*, *23*).

Characterization of the crystal with Raman spectroscopy allows the two phases to be distinguished easily, as the hydrate and anhydrous phases have different Raman signatures for the sulfate (main peak at 990 and 993 $cm^{-1}$, resp.) and water vibrations (present or absent resp.) as shown on Fig. 1C. A more detailed description on the microscopic level will follow.

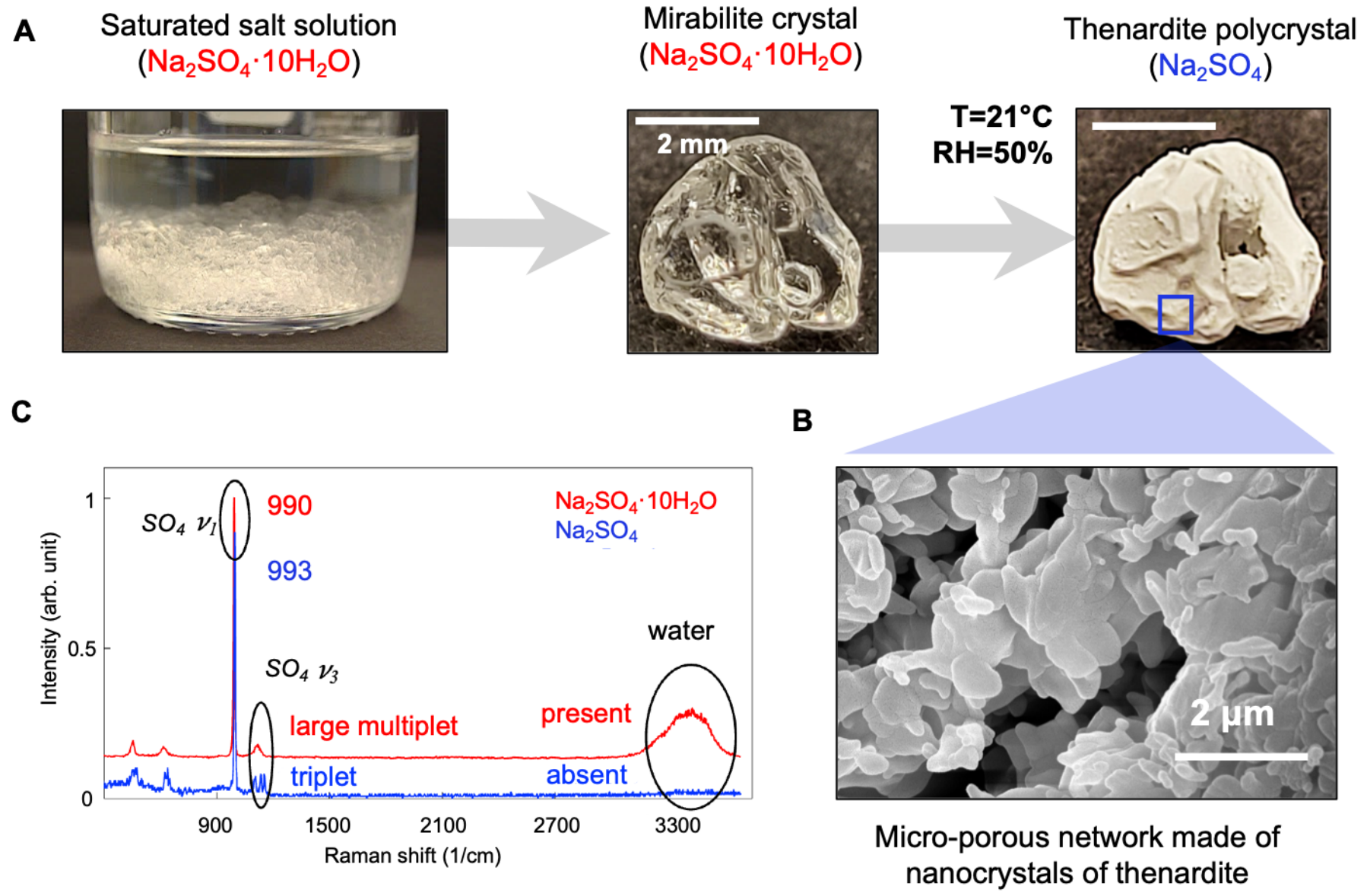

**Fig. 1. Dehydration experiment on mirabilite crystals**. **(A)** Representative crystal of mirabilite during dehydration (scale bar: 2 mm) taken out of a saturated salt solution, turning opaque during dehydration. **(B)** SEM image of a crystal after dehydration, showing a microporous network comprised of nanocrystals of anhydrous phase, thenardite, reproduced with permission from Wijnhorst et al. (*8*). **(C)** Raman spectra of the two phases before (hydrated phase, red) and after (anhydrous phase, blue) dehydration.

To study the impact of RH and temperature on the drying kinetics, we use crystals of mirabilite of similar mass and geometry in dynamic weight loss measurements under a controlled environment. The amount of water loss can be related to the conversion fraction $\alpha$:

$$\alpha = \frac{m_i - m_t}{m_i - m_f} \tag{2}$$

where $m_i$ and $m_f$ the initial and final weight of the crystal respectively, and $m_t$ is the weight at time $t$ of the crystal.

First, a general trend can be observed from the dependency of the conversion fraction $\alpha$ on the RH of the environment (Fig. 2A). For lower RH, *i.e.* larger deviation from the equilibrium relative humidity $RH_{eq}$, dehydration occurs faster as the plateau of reaction completion ($\alpha = 1$) is reached earlier. Second, the functional dependence of $\alpha$ with respect to time is characteristic of the underlying mechanism driving the reaction leading to the transformation between the reactant (hydrated crystal) and the product (anhydrous salt). The shape of the curves shown in Fig. 2A presents a convex curvature for all samples, suggesting that their dehydration is characterized by a similar mechanism. Several physico-geometric kinetic model functions have been developed for solid-solid transformations (*24–27*) that may be used to describe the mechanism. The convex shape of the curve in Fig. 2A shows that dehydration is decelerating throughout. Three mechanisms are compatible with this behavior: *diffusion* control, in which water transport through the reactant or product layer limits the rate; *phase-boundary* control, in which the rate is governed by the advance of the interface between mirabilite and thenardite; and *reaction-order* control, in which the rate depends solely on the fraction of material still unreacted. These three models can be readily distinguished by their integral reaction model, denoted $g(\alpha)$, which scales linearly with time:

$$g(\alpha) = k.t \tag{3}$$

Here, $k$ is the *reaction rate constant* defined by the *Arrhenius law*:

$$k = A.\exp(-E_a/RT) \tag{4}$$

with the frequency pre-factor $A$, $T$ the temperature, $R$ the ideal gas constant, and $E_a$ the activation energy of the rate limiting step. The reaction rate can be expressed as:

$$d\alpha/dt = k.f(\alpha) \tag{5}$$

where $f(\alpha)$ describes the reaction mechanism and is a differential form of the $g(\alpha)$ function.

These equations originate from solid-state kinetics theory and must be corrected for our system in which water vapor is formed during the process. To accommodate this, an extension to the model has been developed (*28*) to replace the *reaction rate constant k* by the *areic growth reactivity* $\phi$, which considers not only the effect of temperature, but also of the water vapor pressure (both self-generated during the dehydration as well as from the environment). After this correction, we obtain equations 6-8 that we can now use to fit our data with the models of diffusion, phase-boundary and reaction-order:

$$d\alpha/dt = \phi.f(\alpha) \tag{6}$$

$$g(\alpha) = \phi.t \tag{7}$$

$$\phi_i = k_i \prod_i [X_i]^{(\beta_i)} - k_i' \prod_i [X_i']^{(\beta_i')} \tag{8}$$

where $k_i$ and $k_i$' is the reaction rate for the direct and reverse reaction resp., $[X_i]$ and $[X_i']$ the concentrations of the reactants and products resp., and $\beta_i$ and $\beta_i'$ the partial reaction order of the reactants and products resp.

The expressions of $g(\alpha)$ for the three models are specified in the Methods section. Following equation 7, a model is considered consistent with the experimental data when $g(\alpha)$ varies linearly with time. No such linear relationship is obtained for the diffusion or the reaction-order models (Fig. 2B-i, iii). By contrast, the phase-boundary model (contracting volume, Fig. 2B-ii) gives a near-linear fit, indicating that the rate-limiting step of dehydration is the reaction at the interface between reactant and product, advancing inward in all three spatial directions. This regime was also inferred by Donkers et al. who studied the dehydration within a sodium sulfate crystal via NMR (*17*).

Subsequently, to identify which specific reaction is responsible for the limitation of the dehydration rate, we extract the value of the areic growth reactivity $\phi$ (fitting parameter) for all experiments at different RH. We can split down the main equation $Na_2SO_4 \cdot 10H_2O$ = $Na_2SO_4$ + 10 $H_2O$ into elementary steps such as the creation of a water vacancy, the diffusion of the water towards the interface, the desorption of the water and the annihilation of the water vacancy. For each of these reactions, we write their reactivity constant and their associated areic growth reactivity (see Supplementary Text and Table S1) and compare it to our experimental data. Knowing that the rate limiting step is located at the interface between the reactant and the product, we find that among those equations for $\phi$, the creation of a water vacancy shows the best match and thus poses as the limiting sub-step (Fig. 2C). The pre-factor in front of the ratio of the partial pressures is equal to 1.36 (inset Fig. 2C) which is slightly larger than the expected value 1 – probably due to the oversimplified model applied here.

In conclusion, macroscopic measurements reveal that the dehydration of sodium sulfate decahydrate is controlled by the progress of the reactant/product interface (phase-boundary), more specifically by the creation of water vacancies at this interface.

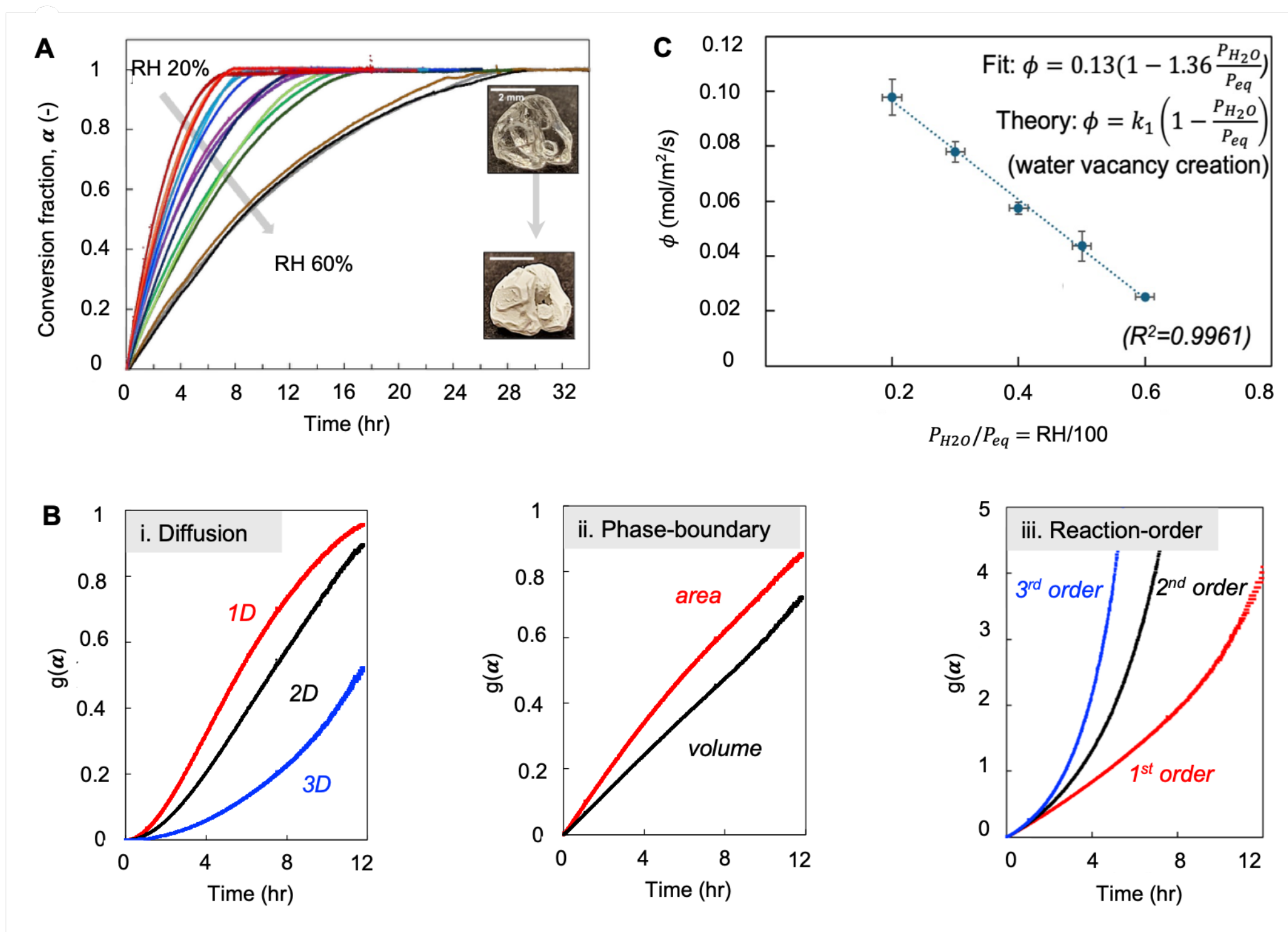

**Fig. 2. Bulk dehydration experiment on mirabilite crystals. (A)** Moisture dependence of the conversion fraction $\alpha$ during dehydration of mirabilite obtained from dynamic weight loss measurements at RH 20% (reds), 30% (blues), 40% (purples), 50% (greens), 60% (blacks) and fixed temperature (20°C ± 1). Three experiments for each RH were conducted and are represented by a different color intensity. The photo and mass of each crystal are given in Fig S1. (B) Integral reaction models $g(\alpha) = \phi \times t$ applied to a crystal dehydrated at RH 40%: i. Diffusion control, ii. Phase Boundary control, iii. Reaction-order control. (C) Moisture dependency of the areic growth reactivity calculated from the fits to our experimental data based on the geometrical contracting volume model applied to all the crystals analyzed in Fig. 2A.

### *Dynamics of dehydration of mirabilite studied by micro-Computed Tomography (micro-CT)*

In parallel, we employ time-lapse X-ray tomography during the dehydration of sodium sulfate decahydrate macro-crystals exposed to RH 25% and 50% to study the macroscale structure of the crystal (Fig. 3). Surprisingly, as dehydration proceeds the crystal shrinks and a layered structure with concentric rings appears inwards through the heart of the crystal (Fig. 3A, Movie S1, Fig. S2). The micro-CT images reveal that the formation of such layers creates a new macroporosity distinct from the microporosity observed by SEM at smaller scale (Fig. 1B). The micro-CT images are processed with a home-developed code, from which we extracted two quantities: the conversion ratio, obtained from the solid volume measured on the X-ray scans over time, and the macroporosity (Fig. 3B). The conversion fraction shows a decelerator behavior over time, in agreement with our weight measurement experiments. At the end of the dehydration, the overall volume reduction (shrinkage) is around 31.5% for the sample at 25% RH and around 36% for the sample at 50% RH. These values are slightly below the theoretical shrinkage 52% calculated from the cell volumes of the hydrate (1460 Å³/unit cell (*29*)) and the anhydrous (701 Å³/unit cell (*30*)) given that the micro-CT only probes only the macroporosity (spacing between those layers) and not the microporosity within those layers (Fig. 1B). The macroscale porosity and layered internal structure provide crucial new insights in the structural deformations that underly the dehydration process.

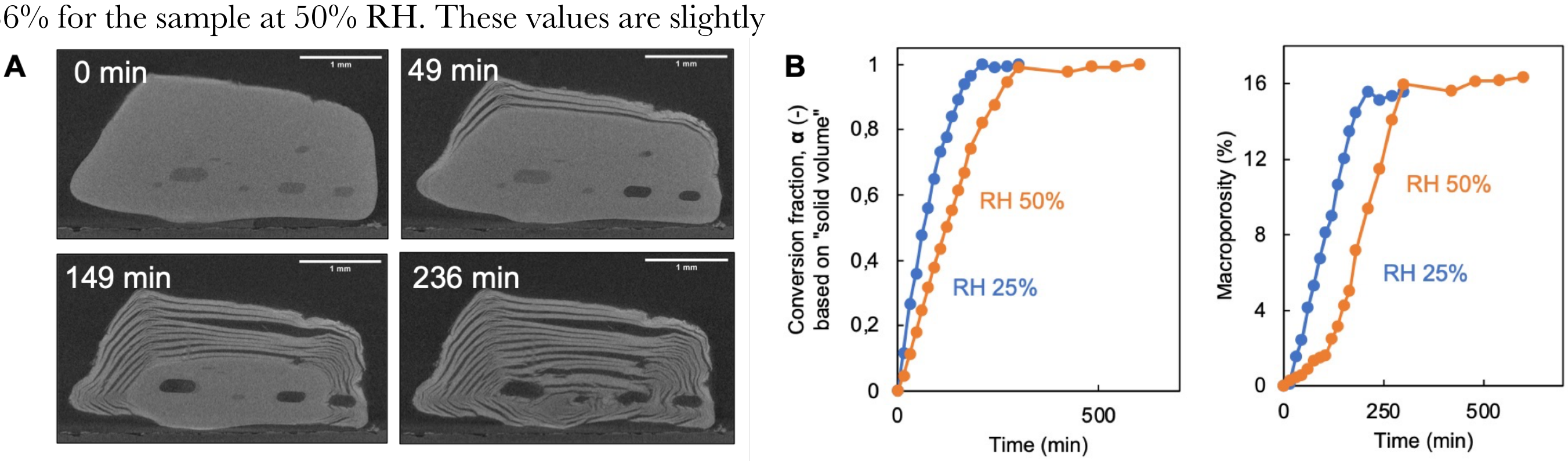

**Fig. 3.** (A) Micro-CT of crystal of sodium sulfate dehydrating at RH 25% (lateral view). (B) Conversion fraction and macroporosity data obtained by treatment of micro-CT scans for two crystals dehydrated at RH 25 and 50%.

## Kinetic study based on microscopic observations

### *Characterization of dehydrated sites*

The macroscale mass loss experiments highlight the rate-limiting step in the dehydration process as well as the impact of RH on the reaction for a long period (time) of dehydration. To gain additional insights in the dehydration mechanism at the very early stage of dehydration, we study the kinetics of the formation of dehydrated zones at microscale at the surface of mirabilite crystal using optical microscopy. At the onset of dehydration, we remarkably observe under the microscope the emergence of black zones that increase in size over time (Fig. 4A), expanding in a circular pattern at a certain rate (movie S2). Those spots grow until the surface of the hydrate is completely covered, without merging; they remain separate from each other similar to grain boundaries often observed in metals (*31*).

Using Raman spectroscopy (Fig. 1C, 4B-F), the local chemical composition inside and outside these spots are characterized. The inner part of the black spots growing in a circular shape at the surface correspond well to thenardite (993 cm$^{-1}$), whereas the outer side, *i.e.* the initial crystalline surface corresponds to mirabilite (990 cm$^{-1}$). Raman mapping along the *x*-*z* plane enables the determination of the 3D geometry of the anhydrous spots as they grow not only laterally but also into the crystal bulk (Fig. 4D, F). Figure 4E shows that the thenardite region grows as a cylinder along z rather than as a sphere. The ratio of growth along *x/y* versus *z* is dependent on the local environment around the dehydration site, as we will show later.

### *Dehydration mechanism for individual dehydration sites*

The experimental data for the conversion factor of two representative dehydration sites (Fig. 4G-H) show an accelerating growth process at the beginning of the dehydration (see Methods for definition of the conversion factor for a single spot). For fitting, since there is no restriction on the growth due to ingestion or coalescence events, we use a power law P2 (*24*, *25*) instead of the Avrami-Erofeyev model. The latter is often used for reactions such as hydration or decomposition reaction by describing solid-state transformations in which the product phase forms by random nucleation followed by growth of the nuclei (*25*).

In our experiments, the initiation and growth of the dehydration spot can be divided in two regimes. In the first regime, *i.e.* initial stage (Fig. 4G-H), the growth is described by a "nucleation" model. After a certain time (for example 6.7 min in Fig 4G and 50 min in Fig 4H), a second regime appears during which the conversion factor is better described by the "phase-boundary" control model. Instead of continuing to grow exponentially, the curve of $\alpha(t)$ flattens out and follows a linear relationship. This is crucially different than the single regime (phase-boundary) behavior that we observe in the macroscale experiments.

The transition from regime 1 (nucleation controlled) to 2 (phase boundary controlled) implies that the *x/y* growth rate slows down compared to the start. This deceleration can be due to different reasons such as: (1) the spot develops rapidly in the 2D plane, but when it enters in the bulk, the dehydration takes longer since the layer of product (thenardite) through which the water vapor now should diffuse to escape in the air becomes larger with the progression of the growth (see Fig. 4I); (2) the presence of and competition with neighboring nucleating sites can also affect the growth rate.

Altogether, our microscale experiments and the application of kinetics theory reveal information on the bulk dehydration that were obscured in standard macroscopic kinetic experiments. Understanding the mechanism of dehydration at the microscale provides information on the dehydration mechanism of the whole crystal (macroscale). Specifically, the kinetic of an individual spot is affected by its surrounding, meaning that our results show that the nucleation and growth of a single dehydrated spot will be reflected in the reaction kinetics of the whole crystal during the dehydration. In fact, the specific spot we are tracking is not considered as a separate reacting entity and the models ("phase-boundary" and "nucleation") at the macroscopic level can still hold for the microscopic case.

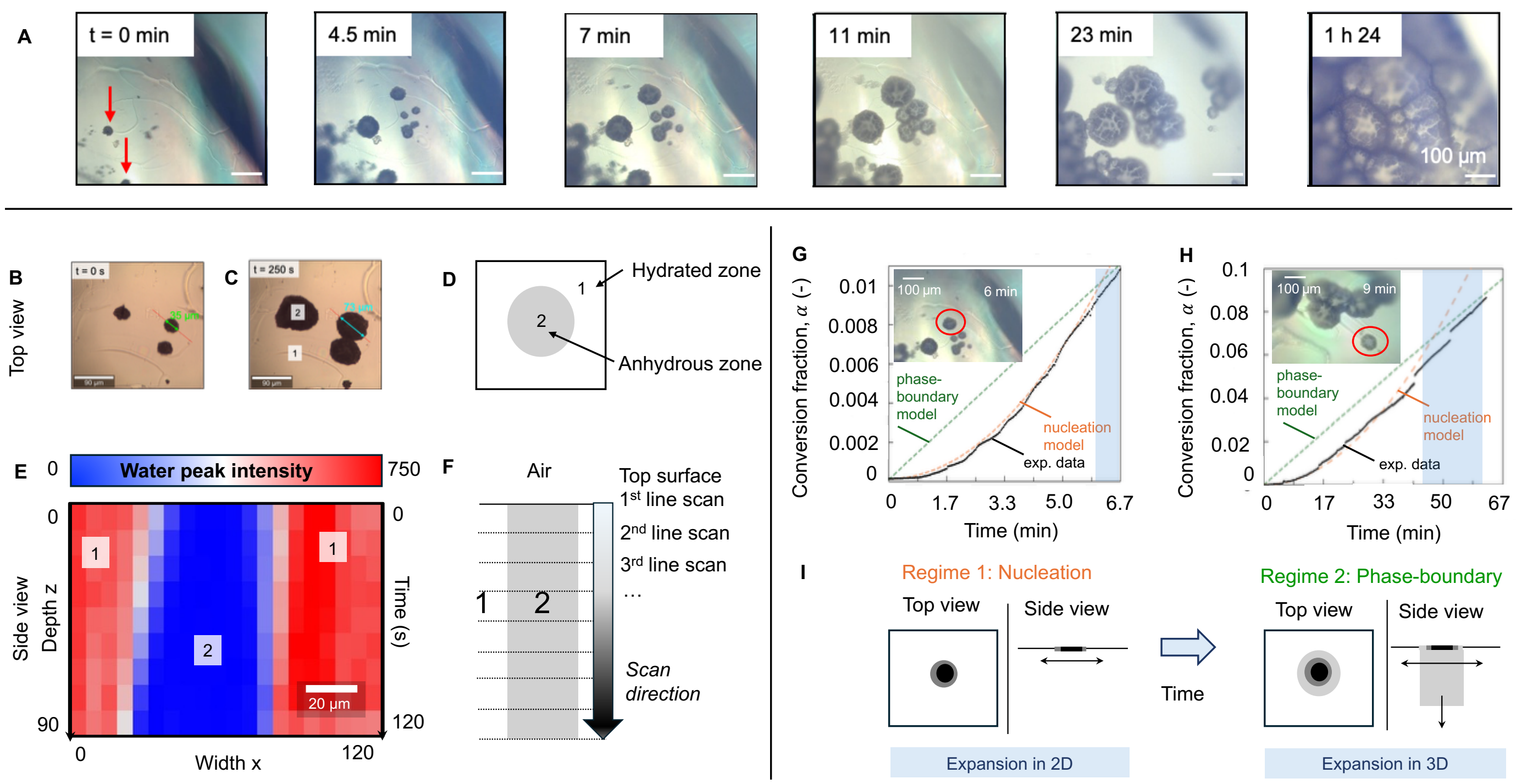

**Fig. 4. (A) Surface characterization of mirabilite crystal during dehydration at RH=60%.** Time evolution of the surface of a sodium sulfate decahydrate crystal upon dehydration. Scale bar: 100 µm. Red arrows highlight nucleation and growth of dehydration zones (*i.e.* formation of the anhydrous phase) of sodium sulfate (thenardite V). **(B-F) Bulk characterization of mirabilite crystal during dehydration at RH=43%.** (B-C) Microscopic top view images of the mirabilite crystal surface during dehydration before Raman scan (t=0 s) and at the end of the scan (t=250 s). (D) Schematic representation of the top surface. (E) Profile view (X-Z) of a growing dehydration zone characterized by mapping in (X-Z) plan using Raman micro-spectroscopy). The color scale describes the intensity of the water peak (position: 3450 $cm^{-1}$, width: 450 $cm^{-1}$) and the labels are zone 1: mirabilite, zone 2: thenardite. The units for depth and width are given in µm. (F) Schematic representation of the experimentally detected profile of a dehydration zone. **(G-H) Growth rate tracking of individual dehydration sites on two different crystals exposed to RH=60%**. The conversion rate $\alpha$ is explained in the Methods part. Our experimental data (black) match with the reaction models used in our earlier macroscopic experiments: 1/ "nucleation" control (orange), 2/ "phase-boundary" control (green). The blue frame marks the change of regime. **(I) Schematic of the 2-step mechanism of nucleation and growth of an individual dehydration zone**.

## Impact of the RH on the microstructure of products

To assess the role of RH in the dehydration process on the resulting microstructure, we compare the microstructure of the products using SEM for mirabilite crystals dehydrated at different RH (Fig. 5A). In all cases, the product is a microporous network composed of an assembly of thenardite nanocrystals, in good agreement with previous works (*8*, *23*). The morphology of these nanocrystals, and the porosity associated with it, nevertheless differ markedly with the relative humidity (RH) of the surrounding environment. At low relative humidity (RH 5%), the polycrystalline structure of thenardite is anisotropic with elongated nanocrystals not very well defined (Fig. 5Ai, ii). At higher relative humidities (Fig. 5Aiii, iv), the assembly is composed of bigger and more spherical nanocrystals, underlining that the growth of the thenardite particles is more isotropic (same speed of growth in the three directions of space) at higher RH. Some papers have already shown how temperature can affect crystal growth anisotropy (*32*–*34*). This effect is known as *thermal roughening* on the atomic scale and leads to rounder crystals (*35*). Although the influence of temperature on the growth anisotropy is well established, our first observations on the influence of the RH motivate the need for further investigations into this largely unexplored parameter.

From these SEM images, we can already infer a difference in the rehydration rate of the products. Thin and small grains with a larger surface area would likely rehydrate faster than larger and spherical particles due to their reduced surface-to-volume ratio. On the other hand, a structure with smaller pores and larger tortuosity would probably take a longer time to rehydrate than a structure with larger pores with less tortuosity as the water vapor would exhibit a reduced mobility through the porous network. The rehydration rate would therefore strongly depend on the competition between the size of the nanocrystals and the properties of the porous polycrystalline, *i.e.* pore size, porosity and tortuosity. Further research should focus on the rehydration mechanism as a function of microscale porosity of these compounds to identify the optimal cycling conditions.

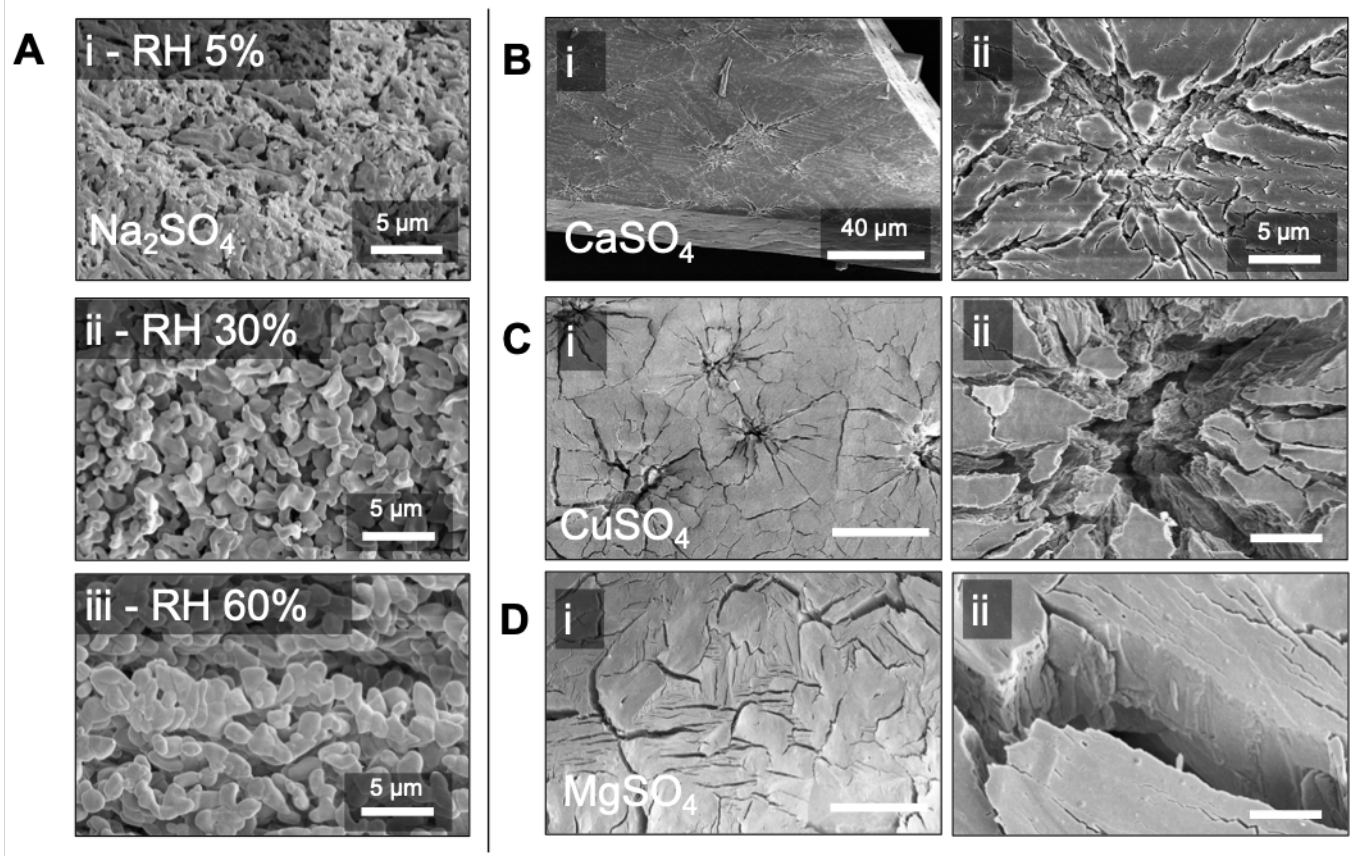

**Fig. 5. SEM images after dehydration of four hydrated crystals: (A) $Na_2SO_4$ at three relative humidities 5% (i), 30% (ii) and 60% (iii) highlighting the difference in microstructure in the product; (B) $CaSO_4$, (C) $CuSO_4$, (D) $MgSO_4$ all dehydrated at 200°C for 16hrs.**

Finally, to generalize our findings we investigate the microstructure of three other sulfate hydrated salts (Table 1) after complete dehydration using SEM. We aim to understand whether the observation of the circular domains (earlier called nucleation sites/spots) are particular to sodium sulfate or is a more universal characteristic of salt hydrates. The three salts have different water content, different volumetric change and crystallographic system transitions between hydrate and anhydrous phase. SEM images (Fig. 5) reveal differences in terms of crack size, crack geometry, crack orientation. $Na_2SO_4$, $CaSO_4$, $CuSO_4$ exhibit similar cracks (nucleation spots), while $MgSO_4$ presents linear cracks instead. It is interesting to note that for $Na_2SO_4$, $CaSO_4$, $CuSO_4$ when transforming from the hydrate phase to the anhydrous phases, there is a change in crystallographic system from low symmetry (monoclinic or triclinic) to a higher symmetry (orthorhombic). Interestingly for $MgSO_4$, there is no change of crystallographic system between the hydrate and the anhydrous product, which then leads to linear cracks (Fig. 5Di, ii). Consequently, these first observations suggest that the difference of crystallographic system between the hydrate and the anhydrous govern the types of cracks on the dehydration product (see discussion below). Moreover, for $CaSO_4$ and $CuSO_4$, the presence of circular nucleation spots clearly presume that their mechanism will be limited by a nucleation process, which is confirmed in the literature (*36*, *37*). On the other hand, $MgSO_4$ doesn't present any circular nucleation spots, nor a high porosity, so we would expect its dehydration mechanism to be limited by diffusion, which in agreement with literature (*38*). The micro-scale structure of a dehydrated product thus provides essential information that allows to infer the limiting mechanism during the dehydration.

## Extension to other sulfate-based hydrated salts

**Table 1: List of hydrated salts studied with relevant information (water molecules lost, volume change between hydrate (H) and anhydrous (A), crystallographic systems). See Figure S3 for Raman spectra for each phase.**

| Compound | $Na_2SO_4$ | $CaSO_4$ | $CuSO_4$ | $MgSO_4$ |
|---|---|---|---|---|
| Crystalline systems | Monoclinic (H) Orthorhombic (A) | Monoclinic (H) Orthorhombic (A) | Triclinic (H) Orthorhombic (A) | Orthorhombic (H) Orthorhombic (A) |
| Volume ratio (H/A) | 2.08 (*29*, *30*) | 1.62 (*39*, *40*) | 1.34 (*41*, *42*) | 3.53 (*43*, *44*) |
| Number water molecules | 10 | 2 | 5 | 7 |
| Crack types | Circular | Circular | Circular | Linear |
| Top surface porosity | High | Low | Low | Low |

## DISCUSSION AND CONCLUSION

Our results highlight that the macroscale dehydration of mirabilite shrouds the underlying microscale dynamics at the level of nucleation and growth of individual dehydrated sites. In fact, the dehydration process consists of 2 phases. In the first phase defined as the "nucleation" control period where the rearrangement of the atoms in the mirabilite structure deprived of water and the formation of the new structure of thenardite is limiting the dehydration process. Our Raman measurements demonstrate that the aspect ratio of the growth sites is non-unitary. As such, we conclude that a spot first grows in the 2D *x-y* plane. As a result, it follows the exponential relationship corresponding to the "nucleation" control model. In the second phase, the growth rate levels off when the dehydration process progresses into the depth of the hydrated crystal. This second regime follows a "phase-boundary" control, *i.e.* where the growth is limited by the reaction happening at the interface between the reactant (hydrated crystal) and the product (dehydrated porous thenardite nanocrystals assembly) – that is the advancement of the interface. The following phase boundary-controlled regime obtained from microscopic observations is therefore meeting the results from the dynamic weight loss (macroscopic experiments).

This interpretation explains why for isolated sites (Fig. 4H) the change of regime occurs later than for sites forming closer to other dehydrated sites (Fig. 4G). An isolated site has more time/space to grow in the x-y direction before reaching a critical value of proximity with the nearest neighbors, which instigates the growth along the z direction. The porous structure of the product revealed by SEM images enables efficient diffusion of the hydration water. Furthermore, the micro computed tomography, revealing the regular layered internal organization of the dehydration product of sodium sulfate, has highlighted the presence of a dual porosity with a shrinkage of around 35%. The first regime of "nucleation" control which is more a surface process at the microscale and starts at the very early stage has been never reported before since in the macroscale study we are following the macro-crystal dehydrating for longer period, rather than that of an individual dehydration site. The macro-crystal is thus averaging out all the dynamics of the individual dehydration sites.

The dehydration rate controlled by the relative humidity of the environment surrounding the hydrated crystal of mirabilite plays an important role in the resulting structure of the product (thenardite). Lower RH favors a fast dehydration resulting to the formation of small thenardite nanocrystals, whereas higher RH leads to slower dehydration rate inducing larger and well-defined nanocrystals of thenardite. We expect that larger crystals would take more time to rehydrate than smaller ones. Further research should be carried to understand the mechanism and kinetics of rehydration based on the structure of the product.

Lastly, our multi-scale kinetic analysis and morphological study on the dehydration dynamics of sodium sulfate can be extended to other hydrated salts. We provide a correlation between the morphology of the anhydrous/hydrated phases observed in SEM, their crystallographic structure and dehydration mechanism. If the transformation between hydrate and anhydrous goes to a more symmetrical system, circular nucleation spots appear. As the unit cell angles between the initial and final phases are different, a rotation must occur because of this induced stress. As the volume of the product is lower than the one of the hydrate, space is created within the final phase, giving rise to the typical nucleation spots visible in Figures 5Bi and Ci. If the crystalline system is the identical between the hydrate and the anhydrous, linear cracks will be observed as no distortion between the crystallographic network is necessary. The presence of nucleation spots on the surface of the dehydrated product aligns with the dehydration mechanism found experimentally for those salts controlled by nucleation process.

In conclusion, we show how multiscale analysis can uncover findings normally hidden under macroscale experiments, but also that microstructure changes between hydrate and anhydrous correlate with the dehydration mechanism. This last point warrants further study to extend it to other family of hydrated salts such as carbonates and oxides, and could result in a new tool for predicting dehydration mechanisms of new hydrated salts interesting for thermal energy storage without resorting to experiments.

## MATERIALS AND METHODS

### Preparation of mirabilite crystals

A solution of sodium sulfate ($Na_2SO_4$, Sigma Aldrich, ≥ 99.0%) is prepared just below the saturation and let open to evaporate the water during several days. Crystals are forming in the bottle and taken out of the solution when they reached a desirable size (1-5 mm radius), dried with a chemwipe to remove the excess of solution surrounding the crystal just before each experiment.

### Preparation of other dehydrated salts

$CuSO_4 \cdot 5H_2O$, $MgSO_4 \cdot 7H_2O$, $CaSO_4 \cdot 2H_2O$, and $K_2CO_3 \cdot 1.5H_2O$ crystals were made by preparing a solution slightly undersaturated from $CuSO_4$ (Merck), $MgSO_4$ (Sigma Aldrich, ≥ 99.0%), $CaSO_4$ (Sigma Aldrich, 99%) and $K_2CO_3$ (Sigma Aldrich, ≥ 99.0%) and let evaporate until crystals form. Crystals were then put in the oven (Memmert) at 200°C over one night to assure complete dehydration. Raman spectra were taken for each of the dehydrated products to check their composition (Fig. S3).

### Raman spectroscopy with confocal microscope

Line scan along z, profile scan along xz and stack scan is investigated by Raman confocal spectroscopy using a WiTec, Alpha 300 R microscope coupled to a CCD camera (Andor, Newton EMCCD, DU970P-BVF-355). The laser wavelength used is 532 nm combined with a diffraction grating of 600 g/mm. For large area scan, we scan in depth (200 µm) with 20 points per line and 20 lines per scan, with an integration time of 0.5 second. The spectra were later processed on the same software of acquisition to build the map presented in Figure 4E, taking the maximum intensity of the water peak at (3450 ± 450) $cm^{-1}$ as criteria for the color scale.

### Dynamic weight loss measurements

We study the dehydration of mirabilite by putting a crystal of mirabilite on a precision scale (0.0001 g) connected to a laptop in order to measure the weight loss over time, due to the evaporation of the hydration water vapor. The set-up is placed in a climatic chamber under controlled humidity. At least three samples of similar mass are recorded for a same RH value.

### X-Ray tomography and image analysis

We performed time-lapse X-ray micro-tomography on one crystal of sodium sulfate decahydrate dehydrated at 25% RH (hydrated crystal sample of 22 mg) and one at 50% RH (of 21 mg). The crystals were scanned using the TESCAN UNITOM XL scanner at the DMEX Centre for X-ray Imaging (Pau, France). Scans were acquired at a voltage of 60 kV and a power of 14 W, with a voxel size of 5 µm. The scanning time for one 360° acquisition was approx. 13 minutes, ensuring a sufficiently fast scan while maintaining a good image quality during the dehydration process. Scans were acquired with an interval of approx. 15 minutes during the first 3 hours of dehydration, followed by an interval of approx. 30 minutes for the following 2 hours, and an interval of approx. 1 hour for subsequent scans. The scans were reconstructed using the PANTHERA software package developed by TESCAN. Although the rate of dehydration is different between those two conditions, the crystal's inner morphological change during the dehydration happens to be similar for both, i.e. layered organised structure.

### Microscopic tracking of dehydration sites forming on the surface of the hydrated crystal during dehydration

To investigate the dehydration mechanism of sodium sulfate decahydrate at the microscopic level, we track the apparition of circular dehydration zones on the surface of mirabilite, by placing a single crystal of mirabilite on a glass substrate under fixed RH and room temperature and following the dehydration under an inverted Leica DM-IRB optical microscope in transmission mode attached to a computer with PixelLink software.

### Determination of the conversion fraction $\alpha$ for the microscopic experiments

The conversion fraction $\alpha$ is proportional to a ratio of mass (mass at time t normalized by the final mass at the end of the reaction), hence to a volume ratio. For the microscope description we use the volume of the dehydration site forming on the surface of the hydrated crystal. At the beginning, we assume a 2D growth of the dehydration site, allowing us to take $\alpha$ as a ratio of surfaces. The surface of the spot over time is measured on each picture. We stop recording the area just before the spot collapses with its neighbours (not to confuse with the end of the dehydration). Thus, for the most isolated sites, we end up with a large area, compared to a site developing closer to a neighbour, which results in a small surface for the last value recorded. As a result, the normalization for $\alpha$ cannot be made with the final area value if we want to compare sites, since we don't know the area of the surface at the end of the dehydration process. To be able to compare growth rates of sites on a same image, we decide to normalize with the full size of the image. This manipulation prevents us to perform a quantitative analysis, *i.e.* to determine any kinetics parameter from the upcoming fits with the reaction model functions. However, we can still use that technique to determine the mechanism limiting the dehydration process as the normalization (with a constant) only shifts the curve $\alpha$ versus time along the y-axis, but the shape of this curve with respect to time remains the same.

Here we are interested in the first instants of dehydration; that's why all sites analysed in this work are tracked since their birth. This is an important point to mention since after some time, a spot can transition to another regime. We studied in total 47 spots, including 14 since their birth. The latter have all shown the presence of the two regimes explained in the main text (regime 1: nucleation, regime 2: phase-boundary).

### Integral model functions g(α) = φ.t used for mechanism determination

For a mechanism limited by *diffusion* (*24*): no nucleation (structural mismatch between reactant and product not too high, no reorganization), easiness of transport of the water molecules within the structure of the reactant/product, through pores to or from the interphase boundary limits the rate of the reaction. Depending on the dimensionality where the diffusion occurs, we have three different expressions:

- 1D: $g(\alpha) = \alpha^2$
- 2D: $g(\alpha) = \big((1-\alpha)\ln(1-\alpha)\big) + \alpha$
- 3D: $g(\alpha) = \left(1-(1-\alpha)^{\frac{1}{3}}\right)^2$

For a mechanism limited by the advance of the interface, called *phase-boundary* control (*24–26*): rapid and dense nucleation, slow growth: $g(\alpha) = 1-(1-\alpha)^{1/n}$ with n=2 for contracting area (nucleation on one surface only) or 3 for contracting volume (nucleation on all surfaces). In that case, the rate limiting step is the reaction at the interface between the reactant and the product.

For a mechanism limited by the *reaction-order* (*25*), the expression of $g(\alpha)$ depends on the order of the reaction (availability of the species):

- Zero-order: $g(\alpha) = \alpha$
- First-order: $g(\alpha) = -\ln(1-\alpha) - 1$
- Second-order: $g(\alpha) = \left[\frac{1}{1-\alpha}\right] - 1$
- Third-order: $g(\alpha) = \frac{1}{2} \times \left[\frac{1}{(1-\alpha)^2} - 1\right]$

For a mechanism limited by *nucleation*: slow and uncrowded nucleation and instantaneous growth of the existing nuclei. The term of "nucleation" means that the rearrangement of atoms between the reactant and the product is the rate limiting step. Here we use the power law P2 to describe the nucleation: $g(\alpha) = \alpha^{1/2}$.

### Scanning Electron Microscopy

The SEM images of the hydrated and anhydrous salts are taken on a Verios 460. Each sample is coated with a gold layer of 30 nm thick using an Agar Sputter Coater.

## Supplementary Materials

**This PDF file includes:**

Table S1
Supplementary Text
Figs. S1 to S3
Legend for movie S1 and S2
Reference

**Other Supplementary Materials for this manuscript include the following:**
Movie S1
Movie S2

## Acknowledgments

H. Derluyn acknowledges the support from the European Research Council (ERC) under the European Union's Horizon 2020 research and innovation programme (grant agreement No 850853). **Funding:** This work was funded by institutional funding of the Institute of Physics at the University of Amsterdam. **Author contributions:** AC, NS and JvdG designed the research. AC conducted the experiments and data analysis. HD performed the X-ray imaging experiments and X-ray image analysis. All authors contributed to the data interpretation and the writing of the manuscript. **Competing interests:** Authors declare that they have no competing interests. **Data and materials availability:** All data are available in the main text or the supplementary materials. Additional data related to this paper may be requested from the authors.

# Supplementary Materials for

## A multi-scale study to unravel the dehydration mechanism of hydrated salts

Anne Claude *et al.*

*Corresponding author. Email: n.shahidzadeh@uva.nl

| Elementary step i | Chemical reaction | Equilibrium constant $K_i$ | Areic growth reactivity $\phi_i$ |
|---|---|---|---|
| (1) Creation of a water vacancy | $\left(H_2O_{H_2O}\right)_{deca} \rightleftharpoons \left(V_{H_2O}\right)_{deca} + \left(H_2O_{i,int}\right)_{anh}$ | $K_1 = \left[\left(V_{H_2O}\right)_{deca}\right]\left[\left(H_2O_{i,int}\right)_{anh}\right]$ | $\phi_1 = k_1(1-\frac{P_{H_2O}}{P_{H_2O,eq}})$ |
| (2) Water molecule moves to the adsorbed layer (interface crystal/air) | $\left(H_2O_{i,int}\right)_{anh} + s \rightleftharpoons H_2O_{ads} - s$ | $K_2 = \frac{[H_2O_{ads}-s]}{\left[\left(H_2O_{i,int}\right)_{anh}\right][s]}$ | $\phi_2 = \frac{k_2}{1+K_1K_2K_4^{1/10}}(1-\frac{P_{H_2O}}{P_{H_2O,eq}})$ |
| (3) Desorption of the water molecule from the adsorbed layer to the air | $H_2O_{ads} - s \rightleftharpoons H_2O_{(g)} + s$ | $K_3 = \frac{P_{H_2O}[s]}{[H_2O_{ads}-s]}$ | $\phi_3 = k_3\frac{K_1K_2K_4^{1/10}}{1+K_1K_2K_4^{1/10}}(1-\frac{P_{H_2O}}{P_{H_2O,eq}})$ |
| (4) Annihilation of the water vacancy | $\left(Na_2SO_{4_{Na_2SO_4}} + 10\,V_{H_2O}\right)_{deca} \rightleftharpoons \left(Na_2SO_{4_{Na_2SO_4}}\right)_{anh}$ | $K_4 = \frac{1}{\left[\left(V_{H_2O}\right)_{deca}\right]^{10}}$ | $\phi_4 = k_4\left(\frac{K_1K_2K_3}{P_{H_2O}}\right)^{10}(1-\left(\frac{P_{H_2O}}{P_{H_2O,eq}}\right)^{10})$ |

**Table S1: Elementary steps of the dehydration reaction of sodium sulfate decahydrate, with their respective expressions of equilibrium constant $K_i$ and areic growth reactivity $\phi_i$**

## Supplementary Text

Demonstration of the expressions for $\phi_1$ and $\phi_2$

The areic growth reactivity was expressed in the *Handbook of Thermal Analysis and Calorimetry - Chapter 5* (*1*) by this general formula:

$$\phi_i = \mathrm{k_i} \prod_i [X_i]^{\beta_i} - \mathrm{k_i}' \prod_i [X_i']^{\beta_i'}$$

with $k_i$ and $k_i$' the reaction rate for the direct and reverse reaction resp., [$X_i$] and [$X_i$'] the concentrations of the reactants and products resp., and $\beta_i$ and $\beta_i$' the partial reaction order of the reactants and products resp.

- For the first elementary mechanism mentioned in Table S1 (creation of water vacancy), we have:

  Knowing that $K_1 = \frac{k_1}{k_{-1}}$:

$$\phi_1 = k_1(1 - \frac{[(V_{H_2O})_{deca}][(H_2O_{i,int})_{anh}]}{K_1})$$

  From reactions (2) and (3):

$$[(H_2O_{i,int})_{anh}] = \frac{P_{H_2O}}{K_2K_3}$$

  From reaction (4):

$$\left[(V_{H_2O})_{deca}\right] = \frac{1}{K_4^{1/10}}$$

  Thus:

$$\phi_1 = k_1\left(1 - \frac{P_{H_2O}}{K_1K_2K_3K_4^{1/10}}\right)$$

  The global equilibrium reaction constant

$$K_{eq} = P_{H_2O}^{1/10} = K_1^{10}K_2^{10}K_3^{10}K_4$$

  Therefore:

$$\boldsymbol{\phi_1 = k_1\left(1 - \frac{P_{H_2O}}{P_{H_2O,eq}}\right)}$$

---

- For the second elementary mechanism:

$$\phi_2 = k_2\left[(H_2O_{i,int})_{anh}\right][s]\left(1 - \frac{[H_2O_{ads} - s]}{K_2\left[(H_2O_{i,int})_{anh}\right][s]}\right)$$

  From reactions (2) and (3):

$$\frac{[H_2O_{ads} - s]}{[s]} = \frac{P_{H_2O}}{K_2K_3}$$

  From reactions (1) and (4):

$$\left[(H_2O_{i,int})_{anh}\right] = \frac{K_1}{[(V_{H_2O})_{deca}]} = K_1K_4^{1/10}$$

  Thus:

$$\phi_2 = k_2\frac{P_{H_2O}}{K_2K_3}[s]\left(1 - \frac{P_{H_2O}}{K_1K_2K_3K_4^{1/10}}\right)$$

  If we assume the conservation of the number of adsorption sites, then we have: $[s] + [H_2O_{ads} - s] = 1$

  From reactions (2) and (3):

$$K_2 = \frac{[H_2O_{ads} - s]}{[(H_2O_{i,int})_{anh}][s]} \Leftrightarrow \frac{[H_2O_{ads} - s]}{[s]} = K_2[(H_2O_{i,int})_{anh}]$$

$$\Leftrightarrow \frac{1}{[s]} - 1 = K_1K_2K_4^{\frac{1}{10}} \Leftrightarrow [s] = \frac{1}{1 + K_1K_2K_4^{1/10}}$$

  Hence:

$$\boldsymbol{\phi_2 = \frac{k_2}{1 + K_1K_2K_4^{1/10}}(1 - \frac{P_{H_2O}}{P_{H_2O,eq}})}$$

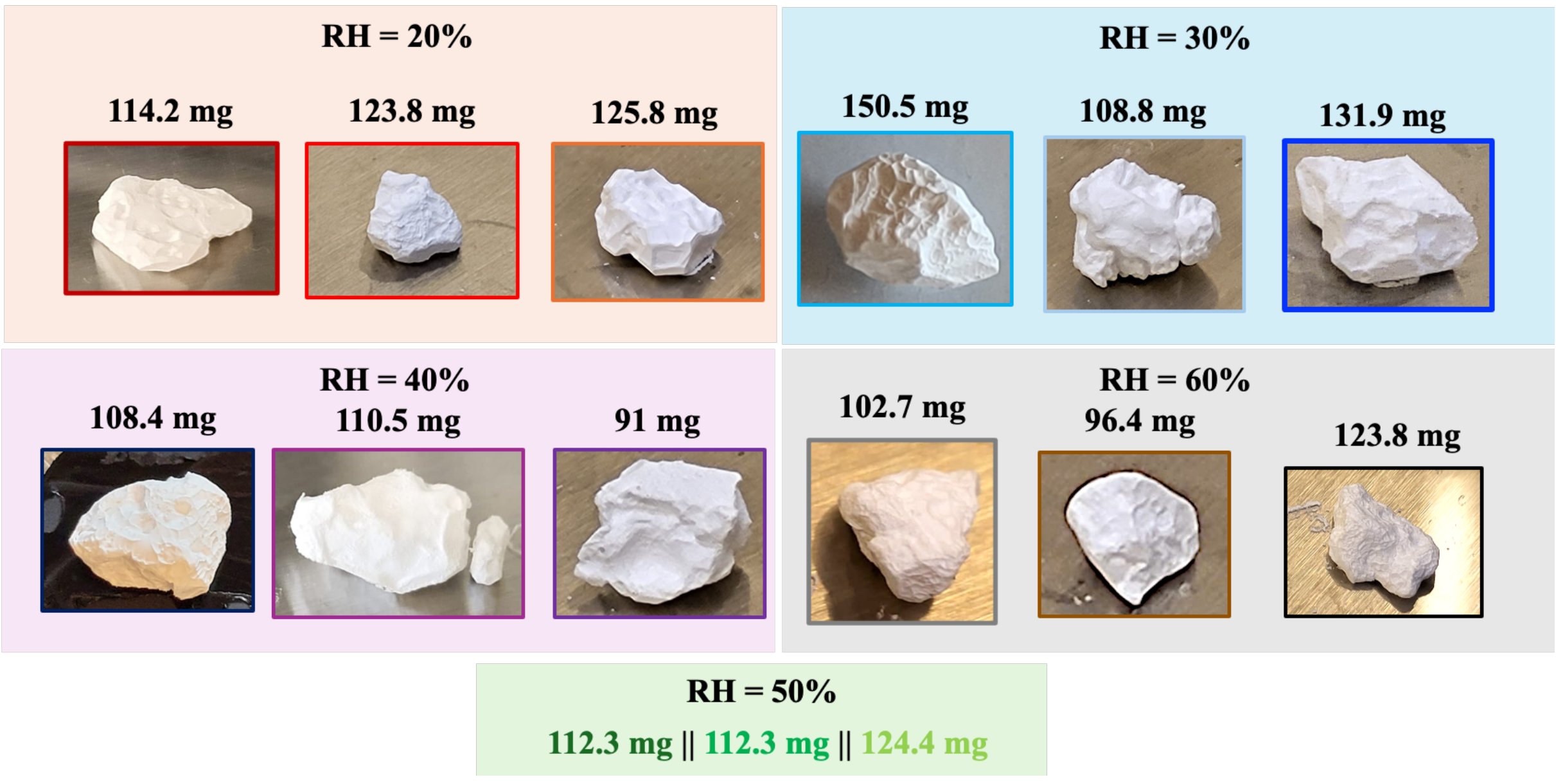

**Fig S1. Photograph and mass of all the bulk crystals studied in Fig. 2A in the main paper**

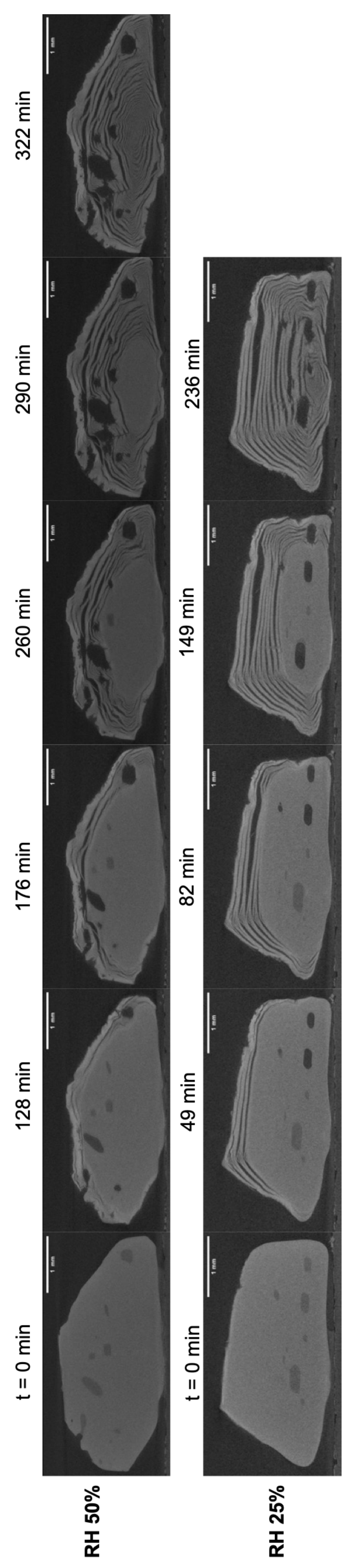

**Fig. S2. MicroCT scans for crystals of sodium sulfate during dehydration at 25% and 50% RH**

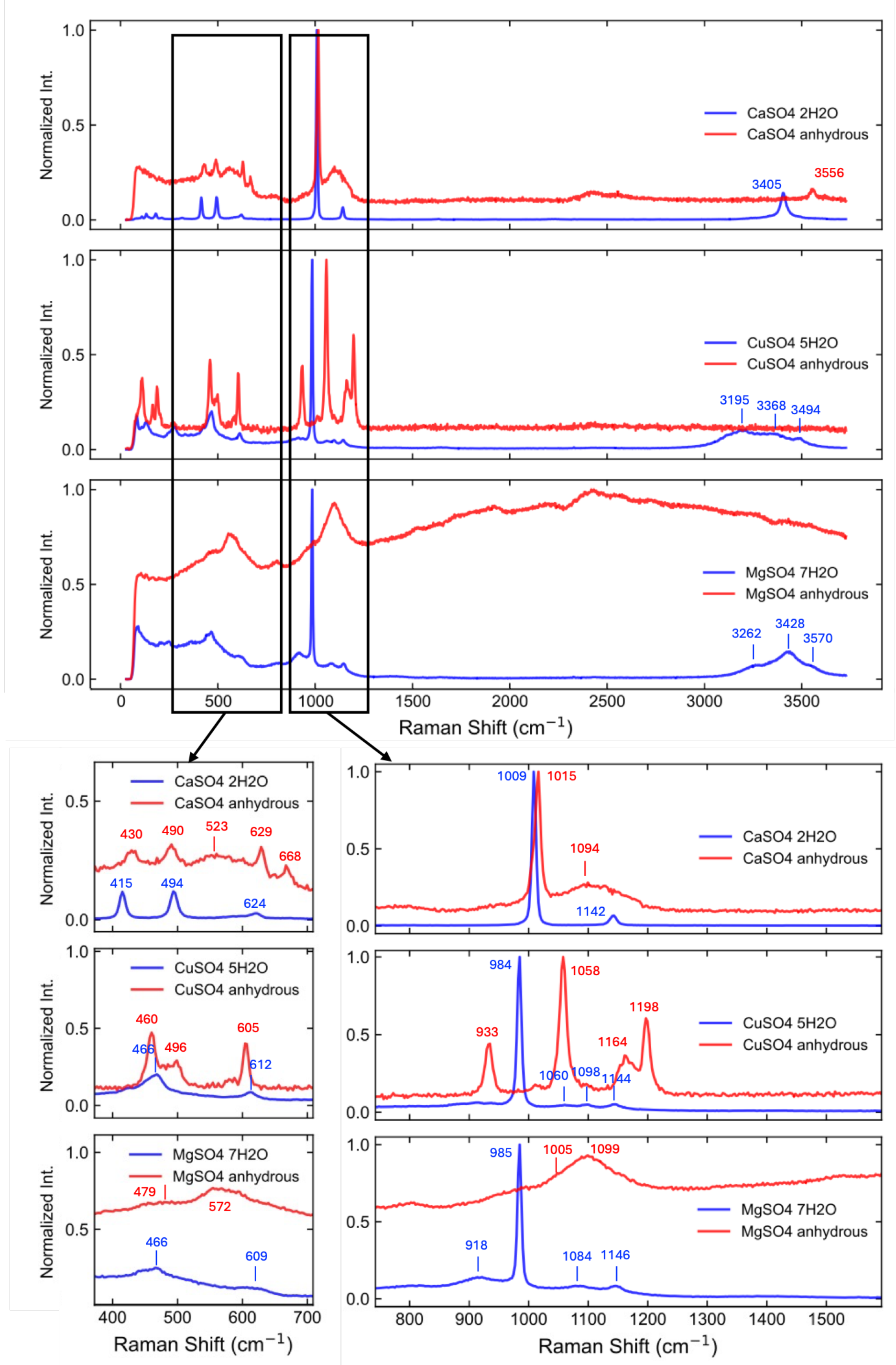

**Fig.S3. Raman Spectra of hydrated and anhydrous salts studied in this paper.**

**Legend movie S1**: Internal restructuration of a crystal of sodium sulfate decahydrate during dehydration at RH25% observed by microCT

**Legend movie S2:** Growing spots of thenardite forming on the surface of mirabilite during dehydration at RH 60%

## Supplementary References